\documentstyle[12pt]{article}
\title{Some Aspects of Trace Anomaly}
\author{\large Martin Schnabl \\[3mm]
\em Nuclear Centre,  \\
\em Faculty of Mathematics and Physics, Charles University, \\
\em V~Hole\v{s}ovi\v{c}k\'ach 2, CZ-18000 Praha 8, Czech Republic}
\date{July 1998}
\begin{document}
\maketitle
\begin{abstract}
We review briefly some well known facts about trace anomaly and
then concentrate on its 'infrared' manifestation. Among other
things we show by means of dispersion relations that dilatations
and translations are conflicting symmetries. We discuss the
occurrence of some subtraction constants in convergent dispersion
relations and illustrate it on the process $H \to \gamma \gamma$
in the electroweak theory.
\end{abstract}

\newcommand{\Tr}{\mathop{\rm Tr}\nolimits}
\newcommand{\re}{\mathop{\rm Re}\nolimits}
\newcommand{\im}{\mathop{\rm Im}\nolimits}
\newcommand{\asympt}{\mathop{\sim}}
\def\bra#1{\langle #1 |}
\def\ket#1{|#1 \rangle}
\def\aver#1{\langle\, #1 \,\rangle}
\let\eps = \varepsilon

\section{Physical aspects}
From the beginning of 1970's we know that dilatation and conformal
symmetries of quantum field theories in flat space-time are in general
broken by quantum effects \cite{CJ,ChE,ACD,CDJ}. The corresponding Noether
currents  thus possess anomalous divergences. This phenomenon has some
important physical consequences. First it explains naturally
fairly large hadronic masses (see e.g. \cite{DGH}).  It also allows one to make interesting
low-energy predictions for various matrix elements \cite{Shifman}.

Another physical case \cite{ChE}, known for a long time, concerns
the asymptotic behaviour of the basic QCD observable $R(s)$ which in the lowest
order is simply given by the quark-parton model (QPM)
\begin{equation}
R(s) \equiv \frac{\sigma_{tot}(e^+ e^- \to \gamma \to hadrons)}{
                  \sigma_{tot}(e^+ e^- \to \gamma \to \mu^+ \mu^-)}
\asympt_{s \to \infty} \sum_i e_i^2
\end{equation}
where $s=q^2$ and $e_i$ are quark charges.
Before QPM was established it was quite unexpected behaviour
since
\begin{eqnarray}
\sigma_{tot}(e^+ e^- \to \mu^+ \mu^-) \sim \frac{1}{s}
\quad {\rm but} \quad
\sigma_{tot}(e^+ e^- \to p \bar p) \sim \frac{1}{s^5}
\end{eqnarray}
Nonvanishing asymptotic limit of $R(s)$ can be understood
\cite{ChE} as a physical consequence of the trace anomaly.
Let us start with the anomalous Ward identity
\begin{equation}\label{LET}
\left(2-p\cdot\frac{\partial}{\partial p} \right) \Pi_{\mu\nu}(p)
= \Delta_{\mu\nu}(p) +
  \frac{R}{6\pi^2}(p_\mu p_\nu - g_{\mu\nu} p^2)
\end{equation}
where
\begin{eqnarray}
\Pi_{\mu\nu}(p) &=&
i \int d^4x e^{ipx} \langle 0| T(J_\mu(x) J_\nu(0)) |0\rangle
\\
\Delta_{\mu\nu}(p) &=&
 \int d^4x d^4y e^{ipy} \langle 0| T(\theta_\alpha^\alpha(x) J_\mu(y)
J_\nu(0)) |0\rangle
\end{eqnarray}
For large momentum $p$ we can neglect the first term on the
right-hand side (RHS) of (\ref{LET}), solve that asymptotic equation and find
the imaginary part
\begin{equation}
\im \Pi_{\mu\nu}(p)
\simeq  \frac{R}{12\pi}(p_\mu p_\nu - g_{\mu\nu} p^2)
\end{equation}
Using unitarity we finally see that the asymptotic ratio $R$ is
nothing else but the trace anomaly
\begin{equation}
\sigma_{tot}(e^{+}e^{-}\to hadrons) =
-\frac{16 \pi^2\alpha^2}{3p^4} \im \Pi_\mu^\mu
\simeq \frac{4\pi \alpha^2}{3p^2} R
\end{equation}

\section{Dispersive derivation of trace anomaly}

Soon after the discovery of axial anomaly based
on some ultraviolet (UV) regularization, Dolgov and Zakharov
\cite{DZ} presented an 'infrared' (IR) derivation.
From their point of view the anomaly appeared as a $\delta(q^2)$
singularity in the imaginary part of some formfactor of the triangle
diagram or as a pole in the real part. Using dispersion relations
and unitarity they completely avoided any UV regularization.

It turns out that this sort of derivation is not confined
only to the axial anomaly but the trace anomaly can obtained
in a similar manner as well \cite{HSch}.
Let us start with the naive dilatation Ward identity
\begin{equation}
\left(2-p\cdot\frac{\partial}{\partial p} \right) \Pi_{\mu\nu}(p)
= \Delta_{\mu\nu}(p)
\end{equation}
Current conservation implies the following form
\begin{eqnarray}
\Pi_{\mu\nu}(p) = \Pi(p^2) (p_\mu p_\nu - p^2 g_{\mu\nu})
\nonumber\\
\Delta_{\mu\nu}(p) = \Delta(p^2) (p_\mu p_\nu - p^2 g_{\mu\nu})
\end{eqnarray}
We can recast the naive Ward identity in terms of these
formfactors
\begin{equation}\label{LETfm}
-2p^2\frac{\partial}{\partial p^2} \Pi(p^2;m^2)
= \Delta(p^2;m^2)
\end{equation}
Let us define $\Pi(p^2)$ and $\Delta(p^2)$ by means of
once subtracted and unsubtracted dispersion relations
respectively
\begin{eqnarray}\label{disprep}
\Pi(p^2) &=& c + \frac{p^2}{\pi} \int_{4m^2}^{\infty}
\frac{\im\Pi(t)}{t(t-p^2)} dt
\nonumber\\
\Delta(p^2) &=& \frac{1}{\pi} \int_{4m^2}^{\infty}
\frac{\im\Delta(t)}{t-p^2} dt
\end{eqnarray}
The imaginary parts are well-defined finite
quantities and should therefore satisfy the canonical Ward identity
\begin{equation}\label{LETim}
-2t\frac{\partial}{\partial t} \im\Pi(t;m^2)
= \im\Delta(t;m^2)
\end{equation}
Plugging the dispersive representation (\ref{disprep}) into the left-hand
side of the Ward identity (\ref{LETfm}), using the naive Ward identity for
imaginary parts (\ref{LETim}) and again (\ref{disprep}) we get
\begin{equation}
-2p^2\frac{\partial}{\partial p^2} \Pi(p^2)
= \Delta(p^2)
- \frac{1}{\pi} \int_{4m^2}^{\infty}
\frac{\im\Delta(t)}{t} dt
\end{equation}
Explicit calculation at one loop level yields
\begin{equation}
\frac{1}{\pi} \int_{4m^2}^{\infty}
\frac{\im\Delta(t)}{t} dt
= -\frac{1}{6\pi} e^2
\end{equation}
The infrared manifestation means that the anomaly
is given by the zero momentum value $ -\Delta(0;m^2)$
and also that
\begin{equation}
\lim_{m\to 0} \frac{1}{t} \im \Delta(t;m^2) = -\frac{e^2}{6\pi}
\delta(t)
\end{equation}
has a singularity in the IR domain.

\section{QED - general case}

An obvious drawback of the preceding derivation is its rather special
kinematical configuration which forbids both the discussion of the
anomalous pair of translation and scale symmetries and the use of
unitarity in different channels.
In analogy with the work \cite{KTV} in scalar $\phi^4$ field theory,
we will study the following Green's functions
\begin{eqnarray}
T_{\alpha\beta\mu\nu}&=&
     \int  dx\,  dy\,  e^{-iqx+iky}
     \bra{0} \theta_{\alpha\beta}(x) J_\mu(y) J_\nu(0) \ket{0}
\nonumber\\
T_{\mu\nu}&=&
     \int  dx\,  dy\,  e^{-iqx+iky}
     \bra{0} \theta_\alpha^\alpha(x) J_\mu(y) J_\nu(0) \ket{0}
\end{eqnarray}
where $q=k+p$. Let us decompose them in the basis of
Bose symmetric tensor structures combined out of $k, p$ and
$g_{\kappa\lambda}$
\begin{eqnarray}
T_{\alpha\beta\mu\nu}&=& F_1(q^2) T_{\alpha\beta\mu\nu}^{(1)} + \dots +
F_{23}(q^2) T_{\alpha\beta\mu\nu}^{(23)}
\nonumber\\
T_{\mu\nu}&=& G_1(q^2) T_{\mu\nu}^{(1)} + \dots +
G_{4}(q^2) T_{\mu\nu}^{(4)}
\end{eqnarray}
assuming for simplicity $k^2=p^2=0$.
The vector, translation and trace Ward identities at the classical level
\begin{eqnarray}
k^\mu T_{\alpha\beta\mu\nu} &=& k_\beta \Pi_{\alpha\nu}(p)
\\
q^\alpha T_{\alpha\beta\mu\nu} &=& k_\beta \Pi_{\mu\nu}(p) +
p_\beta \Pi_{\mu\nu}(k)
\\
T_{\alpha\mu\nu}^\alpha &=& T_{\mu\nu} + \Pi_{\mu\nu}(k) + \Pi_{\mu\nu}(p)
\end{eqnarray}
can be recast in terms of the new formfactors.
Combining these identities we get
\begin{equation}
q^2(F_4 + \frac{1}{2} F_6) = -\frac{1}{2} G_2
\end{equation}
for formfactors standing in front of
$k_\nu p_\mu (k_\alpha k_\beta + p_\alpha p_\beta),\,
 k_\nu p_\mu (k_\alpha p_\beta + p_\alpha k_\beta),\,
 g_{\mu\nu}$ respectively.
Assuming that $G_2(q^2)$ is given by an unsubtracted dispersion
relation we can obtain an anomalous contribution to the RHS
which can be explicitly calculated in the lowest order of QED
\begin{equation}
{\cal A} = \frac{1}{2\pi} \int_{4m^2}^\infty \frac{\im G_2}{t} dt
         = \frac{1}{12\pi^2}
\end{equation}
Now we can assign this anomaly to any of the used symmetries.
If we choose to fix the vector and translation Ward identities we get
anomalous trace identity
\begin{equation}
T_{\alpha\mu\nu}^\alpha = T_{\mu\nu} + \Pi_{\mu\nu}(k) + \Pi_{\mu\nu}(p)
+\frac{e^2}{12\pi^2}(q^2 g_{\mu\nu} -2k_\nu p_\mu)
\end{equation}
which corresponds to the standard result \cite{ACD,CDJ}
\begin{equation}
\theta_\alpha^{\alpha\, anom}=
\frac{e^2}{24\pi^2} F_{\alpha\beta} F^{\alpha\beta}
\end{equation}
Let us note that this anomalous pair can be also studied by various
UV techniques, for the case of point-splitting see \cite{NS}.

\section{Subtraction constants and $H \to \gamma \gamma$ decay}

When we used unsubtracted dispersion relations (DR) for some of the
formfactors we were in fact assuming a vanishing limit for
$k^2 \to \infty$ or $q^2 \to \infty$. This is not always correct.
However in some important cases it is possible to justify such
an assumption proving thus a kind of "No Subtraction Theorem"
without loop integrals calculation; it just suffices to study the
integrands.

These ambiguities due to the subtraction constants can be demonstrated
on a specific physical process of the decay of Higgs boson to
two photons. This one is interesting because the contribution of
heavy particles is given  precisely by the value of the trace anomaly
of the photon field.
Consider just the $W^\pm$ contribution.
The amplitude can be expressed in terms of dimensionless
formfactor $F_W=F_W(m_W^2 / m_H^2)$
\begin{equation}\label{param}
{\cal M}=\frac{\alpha}{2\pi}\frac{1}{v} F_W
(k\!\cdot\! p\, g^{\mu\nu} - p^\mu k^\nu) \eps_\mu^*(k) \eps_\nu^*(p)
\end{equation}

The formfactor $F_W$ is finite and gauge independent. So we
can calculate it e.g. in U-gauge.
It was observed \cite{HS} that although the DR for $F_W$ converge
without any  subtraction we {\it do} need some subtraction constant.
Nevertheless we may note that the off-shell value of $F_W$ which
we use in dispersion relations is {\it not} gauge independent.
Analyzing the integrands of loop integrals we see that the
asymptotic value for $q^2 \to \infty$ is nonvanishing in U-gauge
due to the presence of $1/m_W^2$ terms in $W^\pm$ propagators.
On the other hand we may prove an opposite statement for the
renormalizable gauge  and it can be confirmed by a direct calculation.
Let us remark that direct calculation
in R-gauge is very tedious due to a huge number of diagrams.
The easiest way to get the result is to extract the difference between
the off-shell values in U and R gauges which is given by three diagrams
with unphysical Goldstone bosons.
Thus in a renormalizable gauge we may obtain full trace anomaly
immediately without any subtraction.

\section{Acknowledgments}

I wish to thank Professor J. Ho\v{r}ej\v{s}\'{\i} for guidance throughout the
work and reading the manuscript and J. Novotn\'y for valuable
discussions. This work was partially supported by the grant GA\v{C}R-0506/98.

\end{document}